\documentclass[pre,twocolumn,unsortedaddress,showpacs]{revtex4}

\usepackage{epsfig,rotating}

\begin{document}

\title{Noise-induced inhibitory suppression of malfunction neural oscillators}

\author{C.J.~Tessone$^1$, E.~Ullner$^2$, A.A.~Zaikin$^3$, J.~Kurths$^3$ and R.~Toral$^1$}

\address{$^1$ Institut Mediterrani d'Estudis Avan\c{c}ats (IMEDEA), CSIC-Universitat de les Illes Balears, Ed.~Mateu Orfila - Campus UIB, E07122 Palma de Mallorca, Spain\\
$^2$ Max-Planck-Institut f\"ur Physik komplexer Systeme, N\"othnitzer Stra{\ss}e 38, D01187 Dresden, Germany\\
$^3$ Institut f\"ur Physik, Universit\"at Potsdam, Am Neuen Palais 10, D14469 Potsdam, Germany
}

\date{\today}

\begin{abstract}
Motivated by the aim to find new medical strategies to suppress undesirable neural synchronization  we study the control of oscillations in a system of inhibitory coupled noisy oscillators. Using dynamical properties of inhibition, we find regimes when the malfunction oscillations can be suppressed but the information signal of a certain frequency can be transmitted through the system. The mechanism of this phenomenon is a resonant interplay of noise and the transmission signal provided by certain value of inhibitory coupling.
Analyzing a system of three or four oscillators representing neural clusters, we show that this suppression can be
effectively controlled by coupling and noise amplitudes. 
\end{abstract}

\pacs{05.45.-a, 05.40.Ca }

\maketitle


\section{Introduction}

There are diseases in brain function (specially epilepsy) that involve a large
population of neurons firing pulses synchronously in an oscillatory manner, in
contrast with their typical excitable behavior. It is then important to
understand the mechanisms that, taking advantage of the natural connectivity
between neurons, might help in the suppression of these oscillations related to
the malfunction of the brain. It would also be desirable, besides this
suppression, to restore the transfer functionality such that information could
be transmitted through the neural network.

Several solutions to prevent undesirable neural synchronization have been
proposed. For instance, one could perform a {\it permanent high-frequency 
stimulation} \cite{1991_Benabid,1992_Blond}. To avoid disadvantages of this
strategy (high energy consumption and brain adaptation) the concept of {\it
phase resetting} \cite{1980_Winfree} has been  developed and extended to
populations of interacting oscillators subjected to random forces as {\it
demand-controlled deep brain stimulation techniques}
\cite{1999_Tass_book,2001_Tass_epl}. A promising framework for the development
of future medical methods has been suggested by several recent theoretical
results such as {\it delay feedback control of collective synchrony}
\cite{2004_Rosenblum_prl} or {\it noise-induced excitability}
\cite{2003_Ullner_Zaikin_prl}. In the latter technique not only the
resynchronization of oscillations is achieved but also a restoration of
information transmission is possible. 

The aim of this paper is to present a new method that uses the interplay
between inhibitory coupling and noise, two usual ingredients of living neural
networks, in order to suppress undesirable oscillations but nevertheless being
able to transmit an information signal. More precisely, we investigate in
detail the effects of the coupling, signal frequency and noise intensity on
{\it the oscillation suppression with signal transmission restoration} in a
linear chain of three or four units which combines excitable and
oscillatory neuron models. 

To model the neurons we use the archetype FitzHugh--Nagumo equations
\cite{1961_FitzHugh_biophys} which are a simple example of two-dimensional
excitable dynamics and exhibit a Hopf bifurcation from excitable to oscillatory
behavior as a control parameter is varied. The dynamics of oscillatory and
excitable systems near the Hopf bifurcation attracts large interest because in
this region their sensitivity is greatest and the system is predestined for a
reliable signal response or information exchange. The study of Coherence
Resonance \cite{1996_Gang_prl,1997_Pikovsky_prl} (also named as Stochastic Coherence \cite{jordi}) and Stochastic Resonance in
nonlinear excitable units
\cite{1993_Moss,1994_Wiesenfeld_prl,1999_Anishenko,2004_Lindner_prep} arose a
strong interest on this field. 

It is important to remark here that most of the previous work  was carried out
considering FitzHugh--Nagumo units coupled through its activator variable (the one that
represents the membrane potential), whereas brain neural networks may contain
both activator and inhibitory synaptic connections
\cite{1999_Spiro_j_neurophysiol}. It is known that the inhibitory coupling
between identical oscillators evokes many limit cycles of different periods and
with different phase relations \cite{1991_Volkov_pla2,1994_Volkov2} which are
stable in large regions of the control parameter space and  is usually referred
to as ``de-phasing'' \cite{1995_Han_prl,1999_Postnov} or ``phase-repulsive''
\cite{2001_Balazsi_pre} interaction. De-phasing was shown to be a source of
multi-rhythmicity in different systems
\cite{1992_Sherman,1994_Cymbalyuk,1999_Ruwisch,2002_Volkov_pre}. With neural
noisy elements, a de-phasing interaction of stochastic limit cycles (instead of
deterministic ones) may result in the coexistence of spatiotemporal regimes
selectively sensitive to external signal periods. In such systems, noise plays
at least two roles: first, it stimulates firing of stable elements and,
thereby, their interaction during return excursions; second, it stimulates
transitions between coupling-dependent attractors if the lifetime thereof is
sufficiently long.

In this paper we extend our research on the influence of inhibitory coupling. In contrast to our previous investigation of frequency selective Stochastic Resonance in linear chains of identical excitable FitzHugh--Nagumo models \cite{2003_Volkov_Ullner_pre2,2005_Volkov_Ullner_chaos}, we consider chains of nonidentical oscillators and focus on the influence of the oscillatory units on the dynamics of the whole chain. Further on, we are interested in the suppression of the oscillations and in parallel in an information exchange along the chain across the suppressed elements.

The paper is organized as follows. In the next section II we start with an
investigation of three inhibitory coupled units, each of them representing a
cluster of neural oscillators. A similar architecture may be responsible, for
example, for the activities of neural circuits in a nucleus found in the brains
of songbirds \cite{2002_Laje_prl}. In such circuits  the connection between
different functional units of the brain is mainly due to inhibitory coupling,
whereas the connections within each unit are mainly through the activator
variables.  Since strong activator coupling tends to synchronize the population
of interacting units one can neglect, as a first approximation, that each
functional unit is composed itself of several units, and restrict oneself to a
case in which only one (mean) unit is considered for each region of the brain,
coupled with others through the inhibitor variable. In this architecture the
oscillating element is directly surrounded by inhibitory coupled excitable
elements. To extend the study onto another architecture, in section III, we
analyze a circuit where two coupled oscillating units are connected from both sides with excitable element. Some general conclusions are drawn in the last section IV.

\section{Three non-identical inhibitory coupled FitzHugh--Nagumo}

\begin{figure}[h!t]
\begin{center}
\includegraphics[width=0.45\textwidth]{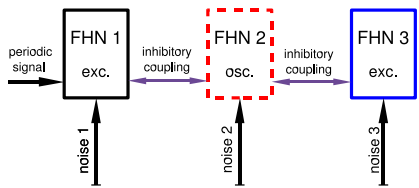}
\end{center}
\caption{Scheme of the setup for the case $N=3$. At both ends there are excitable units, coupled through inhibitor coupling to the middle (oscillatory) element.} \label{fig:setup3}
\end{figure}

We consider a  rather simplistic model with a minimal scheme of connections
that can retain the basic structure of the system we want to mimic: i.e. a
signal injected through a neuron  models the information to be transmitted
along the network with one (or more) malfunction oscillator.  The basic setup
is an open, linear chain where both ends have an excitable unit. The middle
unit represents an ill (oscillatory) unit (Fig.~\ref{fig:setup3}). We want to study whether a periodic, subthreshold signal, acting on the left element can reach the right one, in such a way that no oscillations appear in the middle unit.

The scheme in Fig.~\ref{fig:setup3} corresponds to three FitzHugh--Nagumo 
oscillators coupled through the inhibitory variables:
\begin{eqnarray}
\varepsilon \dot{x}_{1} &=& y_1 - \frac{x_{1}^3}{3} + x_{1} \label{eq:x31} \\
\dot{y}_{1} &=& a_{1} - x_{1}  + \xi_{1}(t) + A_s \cos\left(\omega t \right) + D(y_{2} - y_{1}) \label{eq:y31} \nonumber \\
\varepsilon \dot{x}_{2} &=& y_2 - \frac{x_{2}^3}{3} + x_{2} \label{eq:x32} \\
\dot{y}_{2} &=& a_{2} - x_{2} + \xi_{2}(t) + D(y_{1} - y_{2}) + D(y_{3} - y_{2}) \label{eq:y32} \nonumber \\
\varepsilon \dot{x}_{3} &=& y_3 - \frac{x_{3}^3}{3} + x_{3}  \label{eq:x33} \\
\dot{y}_{3} &=& a_{3} - x_{3} + \xi_{3}(t) + D(y_{2} - y_{3}) \nonumber
\label{eq:y33}
\end{eqnarray}
Where $\omega=2 \pi /{T_s}$, is the frequency of the input signal with period $T_s$. The Gaussian (white) noise sources $\xi_i(t)$ satisfy $\langle \xi_{i}(t)\xi_{j}(t') \rangle = \sigma_a^2 \delta(t-t')\delta_{i,j}$. 

In a neural context, $y_i(t)$ represents the membrane potential of the neuron
and $x_i(t)$ is related to the time-dependent conductance of the potassium
channels in the membrane \cite{1998_Keener_book}. The dynamics of the activator
variable $y_i$ is much faster than that of the inhibitor $x_i$, as indicated by
the small time-scale ratio parameter $\varepsilon$. It is well known that for
$|a_i|>1$ a single unit has a stable fixed point and presents excitable
behavior: small perturbations are followed by a smooth return to the fixed
point, while a perturbation larger than a threshold value induces a return
through a large excursion in phase space (a spike). For $|a_i|<1$, the fixed
point becomes unstable and a stable limit cycle appears. In this regime, the
dynamics consists in a periodic series of spikes. Along this section, we will
consider the fixed parameters: $\varepsilon = 10^{-4}$, $a_{1}=a_3 = 1.01$ and
$a_{2} = 0.99$, such that the two end units are excitable and the middle one,
oscillatory. We have checked that (in the absence of external forcing and
noise) the three units retain their excitable or oscillatory character despite
the coupling amongst them, such that the middle unit spikes periodically and
the two end units display small subthreshold oscillations around the fixed
point.

The issue now is the behavior of these same units when noise and external
forcing are present. We will show that it is possible to have a noise-induced
regime in which the oscillations of the middle unit are suppressed. To
characterize this phenomenon of oscillation suppression, we have
computed \footnote{Most of our results come from a numerical integration of the
previous equations using a stochastic Runge-Kutta type method known as the Heun
algorithm~\cite{1999_smt}.} $N_{s}^{(i)}$, the number of spikes per time unit at
the $i-$th neuron, defined as the number of times the variable $y_i(t)$
surpasses a fixed threshold. $N_{s}^{(i)}$ represents the reciprocal of the
averaged inter-spike time interval.

An important point is whether in this oscillation suppression regime, noise can
help to transmit the information of the subthreshold external signal by a
stochastic resonance mechanism. In order to address this issue, we compute the
linear response, $Q^{(i)}$, of the $i-$th neuron in the chain at the input
frequency $\omega$ \cite{1998_Gammaitoni_reviews,2000_Landa_jphysa}: 
\begin{equation}  
Q^{(i)} =\langle\left|  2y_i(t)\, \hbox{e}^{\imath\, \omega
t}t \right|\rangle, 
\end{equation}  
where $\langle \dots\rangle$ denotes a time average.

\subsection{Oscillation suppression via a noise-induced dynamical trap}

In a previous work~\cite{2003_Volkov_Ullner_pre} it has been shown that in a system of
three FitzHugh--Nagumo units in the oscillatory regime (and in the absence of external
forcing) the inhibitory coupling leads to two coexisting dynamical attractors,
with different natural frequencies. These attractors correspond to an
anti-phase oscillator movement and to the so-called {\it dynamical trap regime}
where the middle oscillator is at rest and the two oscillators at the ends
oscillate in anti-phase. If one now applies a weak external periodic signal to
one of the end units, one can still achieve the suppression of the self-excited
oscillations of the middle unit and, at the same time, achieve a reliable
transmission of the signal, provided the following two conditions hold:
(i) the frequency of the external signal coincides with the natural frequency
of the dynamical trap attractor, and (ii) the noise in the system is near the
optimal one for the desired signal amplification (i.e. stochastic resonance
phenomenon on this attractor). 

\begin{figure}[ht!]
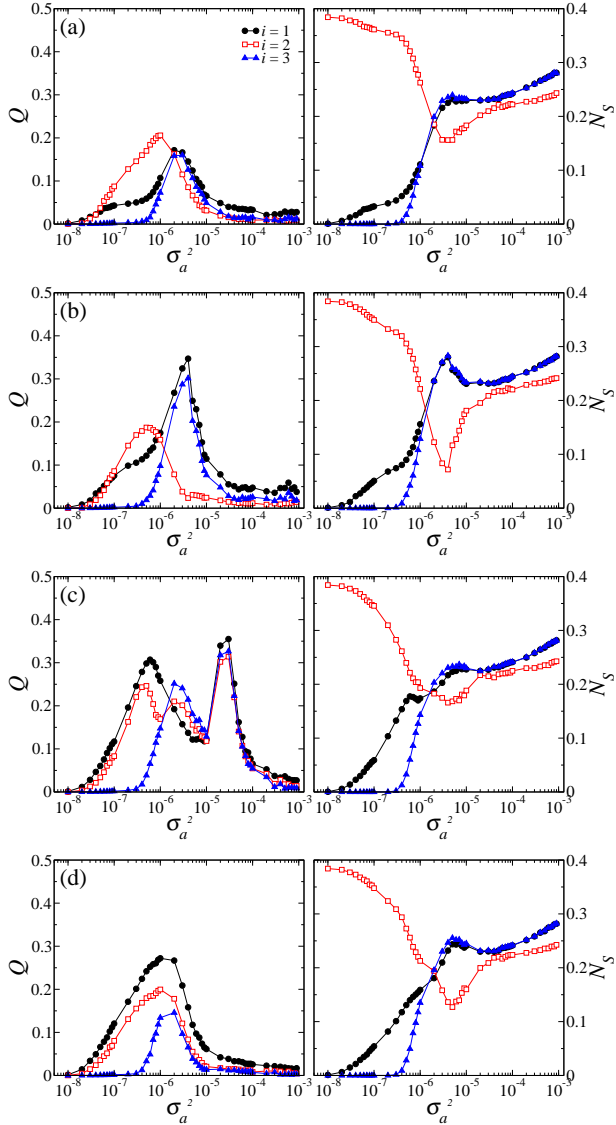

\begin{center}
\includegraphics[width=0.45\textwidth]{fig02a.eps}\\
\includegraphics[width=0.45\textwidth]{fig02b.eps}\\
\includegraphics[width=0.45\textwidth]{fig02c.eps}\\
\includegraphics[width=0.45\textwidth]{fig02d.eps}
\end{center}
\caption{Optimal noise suppresses malfunction oscillations while letting the signal (within a certain range) to be transmitted. This effect occurs due to dynamical trap, supported by inhibitory coupling. (a) non-resonant, $T_{s} = 2.8$;  (b) dynamic trap, $T_{s} = 3.1$; (c) anti-phase resonance, $T_{s} = 4.5$; (d) no resonance, $T_{s} = 6.0$. Other parameters:
$\varepsilon = 10^{-4}$, $a_{1,3} = 1.01$, $a_{2} = 0.99$, $A_s =
0.01$, $D = 0.15$. The left and right columns correspond to the $Q$ and $N_s$ measures.} \label{fig:3QonN_Ts2.8}
\end{figure}

A similar result appears in our system of three coupled units. For very
small noise, the situation is as described at the beginning of the section with
the middle unit oscillating and the end units at rest. As noise increases, one
observes random switches between this state and a dynamical trap regime in
which the middle unit is at rest and the two end units spike in anti-phase.
This effect can be quantified by measuring the number of spikes $N_s^{(i)}$ and
the responses $Q^{(i)}$ as a function of the noise intensity.  As shown in 
Figs.~\ref{fig:3QonN_Ts2.8}, one can distinguish several behaviors depending on
the period of the external forcing.

\vspace{0.5cm}
\noindent {\bf (a)} This is the case where the period of the input signal equals the natural period of an isolated FitzHugh--Nagumo oscillator ($T_s = 2.8$ for $a=0.99$). The noise-induced oscillation suppression described before is apparent in the right panel of this figure, where it is shown that the number of spikes at the middle unit, $N_s^{(2)}$, first decreases as the noise intensity increases. This oscillation suppression is maximum at a value of the noise intensity, $\sigma_a^2 \approx 2 \cdot 10^{-6}$. At noises larger that this value, the number of spikes in the three units are very close to each other. 

In the left panel we plot the response $Q^{(i)}$ of each unit. Note that there is a range of values for the noise intensity for which the middle unit responds to the injected signal most effectively than the end units, as signaled by a higher value of the response $Q^{(2)}$. For increasing noise intensity, beyond the value where the oscillation suppression was maximum, all units have a similar response. 

\vspace{0.5cm}
\noindent {\bf (b)} For an intermediate range of periods $T_s \in [3,3.4]$, we observe that there exists a range of noise intensities ($\sigma_a^2 \in [10^{-6},10^{-5}]$) such that the number of spikes is strongly reduced in the middle oscillatory unit, while the
response to the driving frequency is better than in the oscillatory unit, i.e. this is the manifestation of the {\it dynamic trap regime}. One can clearly see the effective oscillation suppression of the oscillatory middle element (the malfunction neuron, see Fig.~\ref{fig:3QonN_Ts2.8}b, right)
and --despite of this suppression in the middle of the chain-- the reliable information transport from the first to the last neuron model by a large linear response $Q$ in these elements (Fig.~\ref{fig:3QonN_Ts2.8}b, left). The dynamic trap regime includes an anti-phase motion of the first and the last
unit which results in combination with the inhibitory coupling in a suppression of the oscillations of the middle element.

\vspace{0.5cm}
\noindent {\bf (c)} Increasing even further the period, $T_s = 4.5$, the external signal is now in resonance with, and amplifies, the anti-phase motion in which the first and the last
units oscillate in-phase and in anti-phase with the middle one. In this case, another interesting regime appears in the noise range $\sigma_a^2 \in [10^{-5},10^{-4}]$ as observed in the spike
number diagram (right panel of Fig.~\ref{fig:3QonN_Ts2.8}c), where the spike numbers of all three elements coincide nearly, as well as in the linear response plot (left panel), where all oscillators display a very similar linear response $Q$. This anti-phase regime demonstrates a totally different behavior than the dynamic trap regime, case (b)  discussed previously. Note that the anti-phase regime appears for a much larger noise intensity than the dynamic trap regime, hence showing a double selectivity by the input frequency and the noise intensity.

\vspace{0.5cm}
\noindent {\bf (d)} Finally, for much larger period, $T_s = 6.0$, there is no resonance,   Figs.~\ref{fig:3QonN_Ts2.8}d. This can be observed especially at the linear response $Q$ (left panel) which is much smaller than in the resonant cases. Noteworthy, the last element in the chain exhibits a poor signal response.

\subsection{Control of suppression by the coupling strength}

\begin{figure}[ht!]
\begin{center}
\includegraphics[width=0.45\textwidth]{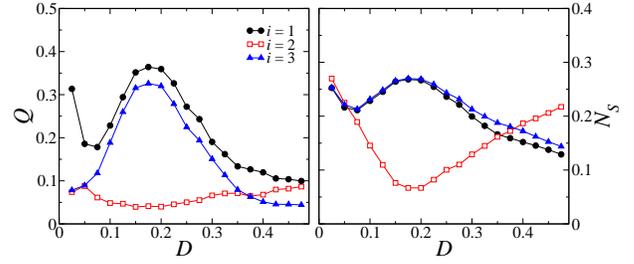}
\end{center}
\caption{Linear response $Q$ (left), and normalized spike number $N_s$ (right), versus inhibitory coupling. The other parameters are: $\varepsilon = 10^{-4}$,
$a_{1,3} = 1.01$, $a_{2} = 0.99$, $A_s = 0.01$, $T_{s} =3.1$ and
$\sigma_a^2 = 3\cdot 10^{-6}$.} \label{fig:3QonD}
\end{figure}

Noise-induced dynamical trap suppression is possible by the existence of a new attractor originated in the inhibitory nature of the coupling. Hence, the coupling intensity, $D$, controls the effectivity of the suppression, as well as the frequency of the attractor. The existence of an optimal value for D is shown in Fig.~\ref{fig:3QonD}, in which we plot the linear response $Q$ and the spike numbers $N_s$ as a function of the coupling in intensity. Starting from the maximum of the resonance curve of the linear response $Q$ on the noise (Fig.~\ref{fig:3QonN_Ts2.8}b right panel), 
it is clearly observed that there is an optimal value of the inhibitory coupling $D$ such that the middle unit is silent (Fig.~\ref{fig:3QonD} right panel), while the first and last units effectively respond to the driving frequency (Fig.~\ref{fig:3QonD} left panel).

\begin{figure}[ht!]
\begin{center}
\includegraphics[width=0.45\textwidth]{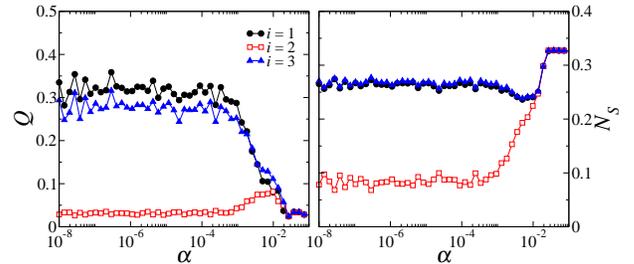}
\end{center}
\caption{The influence of the type of coupling on the linear response $Q$ (left) and the normalized
spike number $N_s$ (right) for three coupled
FitzHugh--Nagumo's. The sliding parameter $\alpha$ shifts the weight of the
diffusion constant $D$ from a pure inhibitory coupling ($\alpha =
0.0$) to a pure activator coupling ($\alpha = 1$). The other parameters are
$\varepsilon = 10^{-4}$, $a_{1,3} = 1.01$, $a_{2} = 0.99$, $\sigma_a^2 = 4\, \cdot 10^{-6}$, $A_s = 0.01$, $T_{s} = 3.1$ and $D = 0.15$.} \label{fig:3percentage}
\end{figure}

Since both types of coupling, inhibitory and activator, can be immanent in neural networks, we have investigated how the suppression can be regulated if we tune the coupling from an activator to an inhibitory one. To do this, we have added activator coupling in the model by interchanging $D$ by
$(1-\alpha)D$ in the equations for the inhibitory variable $y_i$ and inserting the terms $\alpha D\,(x_2-x_1)$ in Eq.~\ref{eq:x31}, $\alpha \,D\,(x_1-x_2)+\alpha D(x_3-x_2)$ in Eq.~\ref{eq:x32}, $\alpha D(x_2-x_3)$ in Eq.~\ref{eq:x33}. These extensions of the model are used only in this section for the calculation of Fig.~\ref{fig:3percentage}. With help of the new sliding parameter $\alpha$ we change the weight of the type coupling from $\alpha = 1$ (pure activator coupling) to $\alpha =  0$ (pure inhibitory coupling).
The results are illustrated in  Fig.~\ref{fig:3percentage}.
We clearly see that increasing the weight of the inhibitory coupling (from right to left) leads an abrupt suppression
of the middle oscillator (Fig.~\ref{fig:3percentage} left) and to a significant joint increase of the linear response $Q$ of the first and third oscillators, but not of the middle one (Fig.~\ref{fig:3percentage} right). Note the logarithmic scaling of the parameter $\alpha$ at the abscissa. We clearly observe, that already a small fraction of activator
coupling (in the order of $1\%$) destroys the dynamic trap regime in the given parameter set.

\section{Model for four non-identical units}

Next we delve into the question of whether larger
chains with more coupled oscillatory units also show  the same phenomenon discussed in the previous section. Although it would seem a rather trivial proposal just enlarge it to a case in which the
system size is $N=4$, the dynamical regimes that arise in such
situation are far from being simple modifications of the results
shown above.

\begin{figure}[ht!]
\begin{center}
\includegraphics[width=0.45\textwidth]{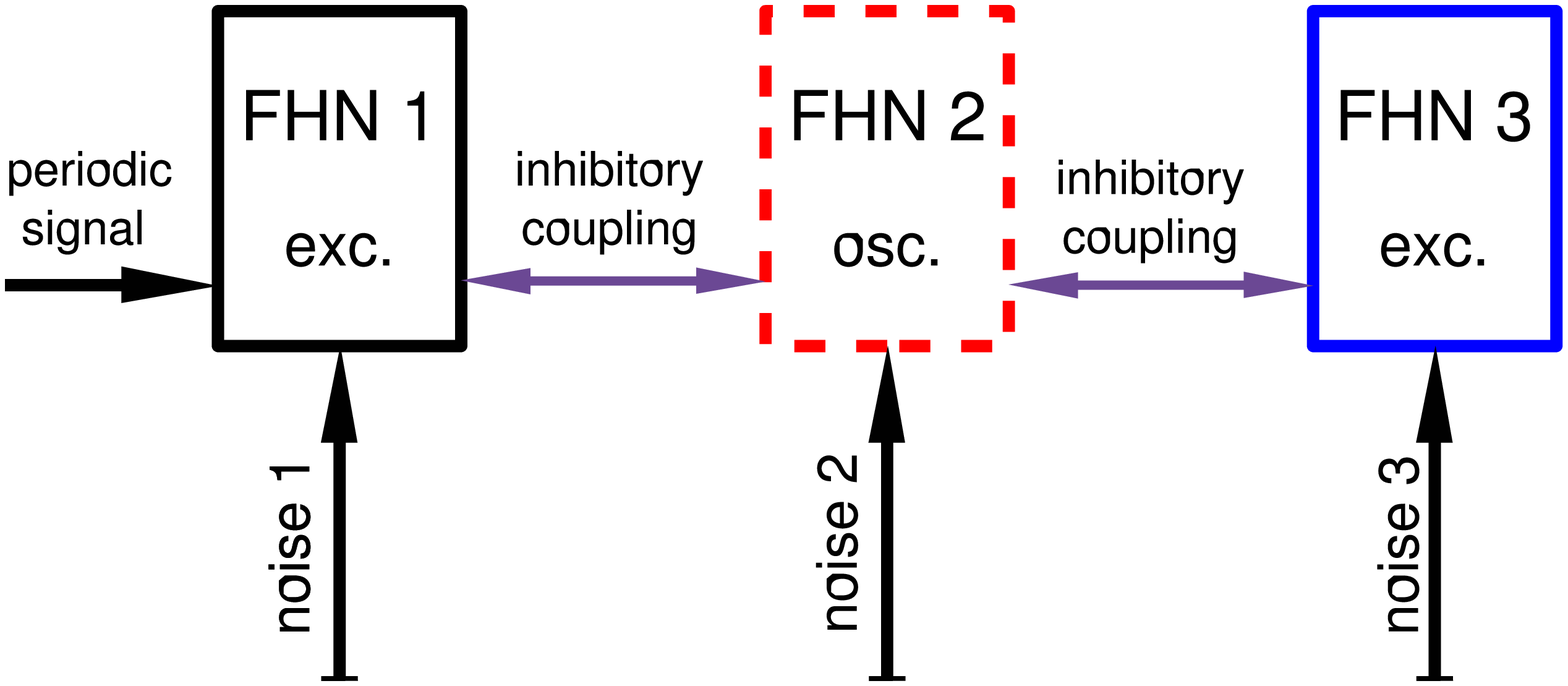}
\end{center}
\caption{Scheme of the setup for the case $N=4$. While at both ends there are excitable units, in the center there are oscillatory ones. The coupling between units of different nature is inhibitory, and the coupling between the oscillatory units is through the activator variable.} \label{fig:setup4}
\end{figure}

We will not consider an enlargement of the excitable ends of the
chain, because it is a well known fact that coupling them through
the activator variable with a strong enough bind, will result in an
entrainment of such subchain and the dynamical evolution of such
units will be effectively that of one oscillator. Then, the most
interesting question arises from the enlargement of the middle
part, that is composed by oscillatory (malfunctioning) units. So, we add to the scheme of three elements (Fig.~\ref{fig:setup3}) an oscillatory element in the middle position and couple it by an activator coupling with the other (identical) oscillatory element and with an inhibitory coupling to the
adjacent excitable element (Fig.~\ref{fig:setup4}).

The mathematical description of the scheme in Fig.~\ref{fig:setup4} is given in  Eqs.~\ref{eq:x41}-\ref{eq:x44}. The two oscillatory units (malfunction neurons) are placed at the middle position and are both coupled to their adjacent
excitable one by an inhibitory coupling as in the chain of three elements, whereas an activator coupling is set between them. As in the previous section, independent additive white noises act on the units and an external, subthreshold, periodic signal drives only the first
element:
\begin{eqnarray}
\varepsilon \, \dot{x}_1 &=& y_1 - \frac{x_1^3}{3} + x_1 \label{eq:x41}\\
\dot{y}_1 &=& a_1-x_{1}+ \xi_{1}(t) + A_s \sin \left( \omega t \right) + D (y_{2} - y_{1}) \label{eq:y41} \nonumber \\
\varepsilon \, \dot{x}_2 &=& y_2 - \frac{x_2^3}{3} + x_2 + C (x_{3} - x_{2}) \label{eq:x42}\\
\dot{y}_2 &=& a_2-x_{2}+ \xi_{2}(t) + D (y_{1} - y_{2}) \label{eq:y42} \nonumber \\
\varepsilon \, \dot{x}_3 &=& y_3 - \frac{x_3^3}{3} + x_3 + C (x_{2} - x_{3}) \label{eq:x43}\\
\dot{y}_3 &=& a_3-x_{3}+ \xi_{3}(t) + D (y_{4} - y_{3}) \label{eq:y43} \nonumber \\
\varepsilon \, \dot{x}_4 &=& y_4 - \frac{x_4^3}{3} + x_4  \label{eq:x44}\\
\dot{y}_4 &=& a_4-x_{4}+ \xi_{4}(t) + D (y_{3} - y_{4}) \label{eq:y44} \nonumber
\end{eqnarray}

We will fix along the following simulations the parameters:
$a_{2,3} =0.99 $ (oscillatory regime), $a_{1,4} = 1.01 $ (excitable
regime) and the signal intensity $A_s = 0.01$ (subthreshold).

We are interested in the signal penetration along the chain from the first to the last element as a function of the signal period and the noise intensity. In order to investigate whether the same phenomenon appears in this chain, two different cases are considered: first, we take the optimal parameters from the case $N=3$ and make the coupling between the oscillatory units strong enough such that they become entrained. In the second case, we use a weaker activator coupling.

\subsection{Strong inter-oscillatory coupling}

Let us focus first on a regime of strong coupling among the oscillatory units. We use the following set of parameters $\varepsilon = 10^{-4}$, $a_{1,4} = 1.01$, $a_{2,3} = 0.99$, 
$C =0.80 $ and $D =0.22$. In this case, and without an external periodic signal ($A_s = 0.0$) injected nor noise ($\sigma_a^2 =0.0$), the analysis of the power spectrum exhibits that the natural period of the system is $T_{nat} \approx 2.67 $. 
The oscillatory units
exhibit their periodic oscillations at their natural frequency.
The excitable units, at their time, show only subthreshold
oscillations at the natural frequency of the oscillatory units.

\begin{figure}[ht!]
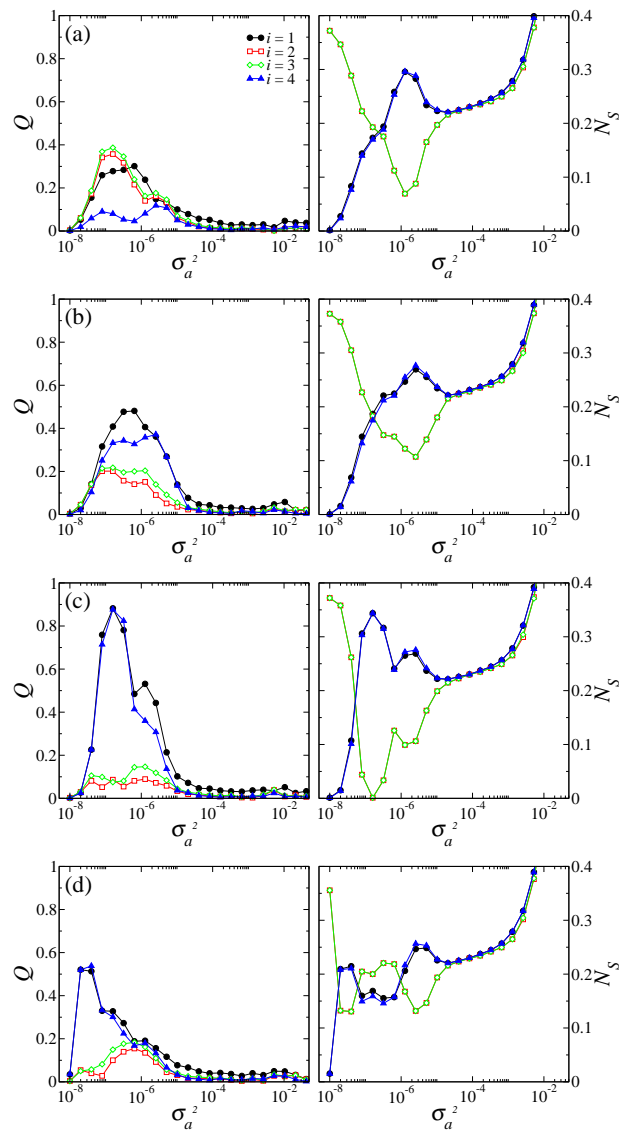

\begin{center}
\includegraphics[width=0.45\textwidth]{fig06a.eps}\\
\includegraphics[width=0.45\textwidth]{fig06b.eps}\\
\includegraphics[width=0.45\textwidth]{fig06c.eps}\\
\includegraphics[width=0.45\textwidth]{fig06d.eps}
\end{center}
\caption{Linear response $Q$ (left column) and $N_s$ (normalized spike number), in the right column, versus noise intensity for a chain of four oscillators. (a)  $T_s = 2.61$; (b) $T_s = 2.8$; (c) $T_s = 2.9$; (d) $T_s = 3.1$; Other parameters: $\varepsilon = 10^{-4}$, $a_{1,4} = 1.01$, $a_{2,3} = 0.99$, $A_s = 0.01$. The couplings are, $C =0.80 $ and  $D =0.22$ (strong coupling regime).  and $T_s = 2.61$.}
\label{fig:4QonNsc_Ts2.61}
\end{figure}

In the presence of an external signal Fig.~\ref{fig:4QonNsc_Ts2.61} illustrates the normalized spike number and the linear response $Q$ as a function of the noise intensity $\sigma_a^2$ for different driving periods $T_s$. Fig.~\ref{fig:4QonNsc_Ts2.61}a depicts the results when the system is subjected both to noise and external signal and the signal period $T_s = 2.61$ is slightly below the natural period. It is observed (as in the $N=3$ case) that now the oscillatory units respond but not the excitable ones.

Increasing $T_s$ well over the natural frequency, e.g. $T_s=2.8$ or $2.9$ (Figs.~\ref{fig:4QonNsc_Ts2.61}b or \ref{fig:4QonNsc_Ts2.61}c), the dynamic trap regime appears. It is important to emphasize that the quality of the signal transmission to the last unit is enhanced with respect to the $N=3$ case (compare Figs.~\ref{fig:3QonN_Ts2.8} and Figs.~\ref{fig:4QonNsc_Ts2.61}).

\begin{figure}
\includegraphics[width=0.3\textwidth]{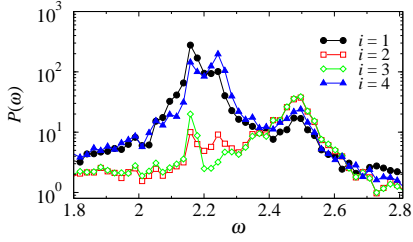}
\caption{Power spectrum of a four oscillator system for the parameters: $\varepsilon = 10^{-4}$,
$a_{1,4} = 1.01$, $a_{2,3} = 0.99$, $A_s = 0.01$, $T_s = 2.9$, $C =0.80 $, $D
=0.22 $, and $\sigma_a^2 =2.56 \cdot 10^{-6}$.}
\label{fig:psd4-2}
\end{figure}

An interesting phenomenon occurs for $T_s=2.9$ (Fig.~\ref{fig:4QonNsc_Ts2.61}c) where there are two
well-differenced situations of dynamics trap like regimes. First, for
very low noise intensities ($\sigma_a^2 \approx 10^{-7}$) there is an almost perfect suppression
of the oscillations and at the same time a perfect signal
transmission which is the result of the desired dynamic trap regime. 
There is a then a secondary oscillation
suppression regime at $\sigma_a^2 \approx 2.5\cdot 10^{-6}$, at which the signal is not transmitted with
the same fidelity as compared to the case at about $\sigma_a^2 \approx 10^{-7}$. What is happening in the second regime is that the last
unit is oscillating, neither at the driving frequency, nor at the
natural one of the middle oscillators, but at another one. Fig.~\ref{fig:psd4-2} shows the power spectrum for such secondary regime in the interesting frequency range around driving and resonance frequency. Let us consider the particular case of the fourth oscillator. It is subject to two different 
signals, one of them with the natural frequency of the third unit, and one with 
the external driving frequency. It is not trivial how this two signals 
interact in order to produce this unit's response, but it has been 
demonstrated that in non-linear systems \cite{2002_Chialvo_pre} subjected to two signals, the response may appear at neither any of the driving ones. Nevertheless these facts, the important footprint of this secondary regime is the low response of the last unit to the driving frequency.

One can clearly see in Fig.~\ref{fig:psd4-2} three peaks in the frequency range $\omega \in [1.8,2.8]$ in the system output. The first and highest peak at $\omega \approx 2.16$ is well pronounced only for the first and driven oscillator and corresponds to a period $T \approx 2.9$, equal
to the driving period $T_s$, i.e. only the driven oscillator exhibits a good response to the signal. The second peak, very close to the first one, at $\omega \approx 2.245$ ($T \approx 2.8$) is displayed mainly by the last unit. The third peak at  $\omega \approx 2.49$ ($T \approx 2.52$) can be found in
all elements with nearly equal hight. The third peak has a very small influence on the total responses of the
system (note the logarithmic scale) and it is produced by small sub-threshold oscillations. The corresponding time series is as follows: the first (driven) oscillator shows a reliable spiking behavior with a period equal to the driving period, while the two middle oscillators are mostly silent and
the last oscillator spikes with a slightly reduced period of $T_s = 2.8$ (so leading to the
difference in the linear response at the signal frequency $Q$ between the first and last FitzHugh--Nagumo in this regime). Due to the small difference in the periodicity, there is no phase locking in this regime and a continuous phase slip between the first and last unit appears.  If the phase difference is large
enough, the chain switches to an anti-phase regime, i.e. the otherwise silent middle oscillators spike in anti-phase to their excitable neighbors. This transition to the anti-phase attractor induces a delay of the last unit compared to the first one. This anti-phase regime is unstable at the considered
parameter set and the chain switches back to the previous attractor with the silent middle elements and the phase slip between the first and last one. The interruption of this long-life attractor by the unstable anti-phase attractor results in a nearly equal spike number of the first and last
unit. Therefore, the interesting behavior in Figs.~\ref{fig:4QonNsc_Ts2.61}c at a noise intensity $\sigma_a^2 \approx 2.5 \cdot 10^{-6}$ is caused by a regime which is only similar to the dynamic trap regime, but is not exactly the desired dynamic trap and hence does not provide a reliable
information transport.

\begin{figure}[ht!]
\begin{center}
\includegraphics[width=0.45\textwidth]{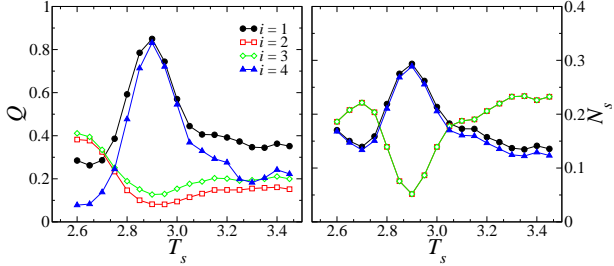}
\end{center}
\caption{Linear response $Q$ and normalized spike number $N_s$ versus time periodicity $T_s$. The system is composed by four units in the strong-coupling regime and the rest of parameters, are:
$\varepsilon = 10^{-4}$, $a_{1,4} = 1.01$, $a_{2,3} = 0.99$, $A_s = 0.01$, $C =0.80 $, $D =0.22$, and $\sigma_a^2=2\cdot 10^{-7}$. }
\label{4QonTsc}
\end{figure}

As a summary of this section, it could be said that the dynamic trap regime still occurs, but
in a narrower region of the driving period ($T_s \in [2.8,3.0]$, Fig.~\ref{4QonTsc}) than in the case of $N=3$ ($T_s \in [3.0,3.4]$).

\begin{figure}[ht!]
\begin{center}
\includegraphics[width=0.45\textwidth]{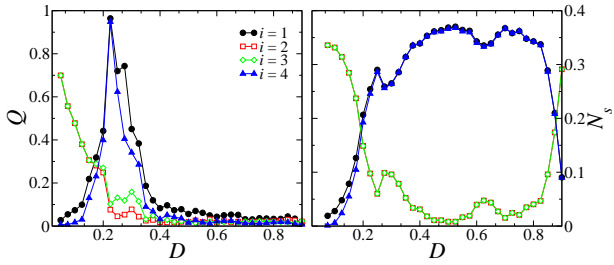}
\end{center}
\caption{Linear response $Q$ and normalized spike number $N_s$ as a function of the inhibitory coupling strength $D$. The system is composed by four units in the strong-coupling regime and the rest of parameters, are: $\varepsilon = 10^{-4}$, $a_{1,4} = 1.01$, $a_{2,3} = 0.99$, $A_s = 0.01$, $C =0.80 $,
  $\sigma_a^2 = 2\cdot 10^{-7}$ and $T_s = 2.9$.} \label{4QonCsc}
\end{figure}

Finally, Fig.~\ref{4QonCsc} yields that there is also a range of inhibitory coupling $D$ such that this phenomenon happens. This resonance like behavior with respect to the inhibitory coupling strength is caused by the influence of this parameter on the resonance frequency of the dynamic trap regime. This figure shows the existence of a maximum (located at a coupling $D \approx  0.25$) in the response as a function of this parameter.

\subsection{Intermediate inter-oscillatory coupling}

\begin{figure}[ht!]
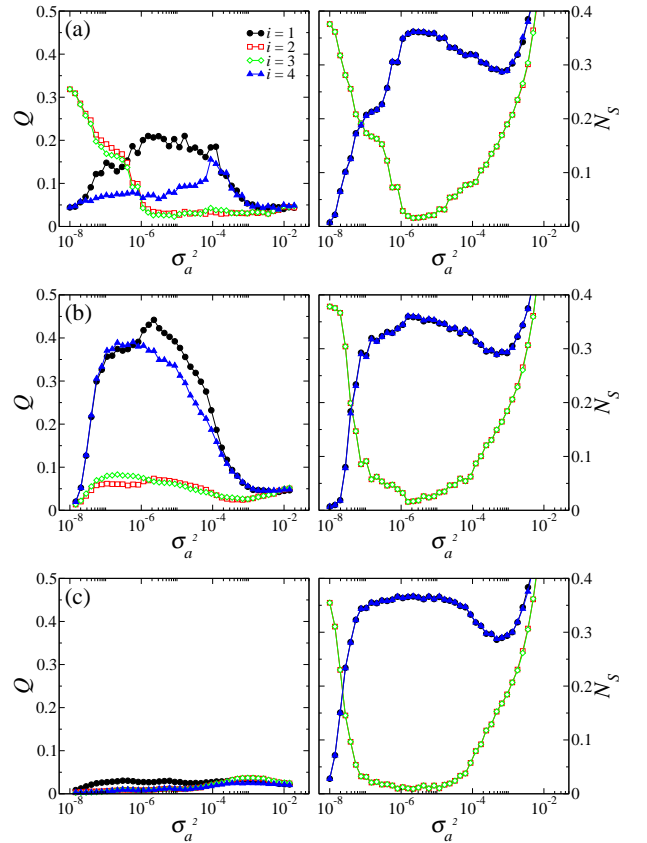

\begin{center}
\includegraphics[width=0.45\textwidth]{fig10a.eps}\\
\includegraphics[width=0.45\textwidth]{fig10b.eps}\\
\includegraphics[width=0.45\textwidth]{fig10c.eps}
\end{center}
\caption{Linear response $Q$ and normalized spike number $N_s$ (left and right columns, respectively) versus noise intensity. The time periodicities, are (a) $T_s = 2.55$; (b) $T_s = 2.61$; (c) $T_s = 5.2$. The other parameters are: $\varepsilon = 10^{-4}$,
$a_{1,4} = 1.01$, $a_{2,3} = 0.99$, $A_s = 0.01$, $C =0.20 $, and $D
=0.50$. } \label{fig:4QonNwc_Ts2.55}
\end{figure}

We also found another kind of dynamic regime in this model with a smaller activator coupling. The analysis of the power spectrum in the absence of noise and external signal shows that  the oscillatory units exhibit their periodic oscillations at their natural period $T_{nat} \approx 2.54$. The excitable units, at their
time, generate only subthreshold oscillations at the natural frequency of the oscillatory units. Note that the natural frequency is shifted from the previous case of a strong inter-oscillatory coupling ($T_{nat} \approx 2.67$). In this case, however, the dependence with $T_s$ of the linear response $Q$ curves and the oscillation suppression is quite different from the previous case. Even for slightly
detuned input signals $T_s = 2.55$ a strong dynamic trap arises in the system (Fig.~\ref{fig:4QonNwc_Ts2.55}a). This oscillation suppression mechanism is very robust over a wide range of driving periods $T_s$  (Figs.~\ref{fig:4QonNwc_Ts2.55}a-c), whereas a reliable signal transmission along the chain can be
observed only in a much narrower range of the driving period, $T_s \in [2.6,2.65]$ (Fig.~\ref{fig:4QonNwc_Ts2.55}b).

The oscillation suppression here is really robust, showing that the middle units do not spike for very large periods of time. The phenomenon is also robust to changes of almost four decades in the  noise intensity.

Note that the curves for the number of spikes show an exact coincidence between
the first and the last units (i.e. the excitable ones), although such a perfect matching does not occur for the linear response $Q$. The first and the last units fire at the same rate (same normalized spike number $N_s$), but they are not phase locked, i.e. there is a random phase slip.  When the difference in phase between these
two excitable units is large enough, this dynamic regime destabilizes and a regime in
which there is an in-phase motion of excitable units, and (in anti-phase) spikes of the
oscillatory units appears. But this last dynamic regime is unstable and rapidly falls
to the previous one. It is interesting that the matching in the number of spikes occurs
in the dynamic trap regime, i.e. that the sub-threshold dynamics of the oscillatory
units is sufficient to carry information from one end of the chain to the other one.

\begin{figure}[ht!]
\begin{center}
\includegraphics[width=0.45\textwidth]{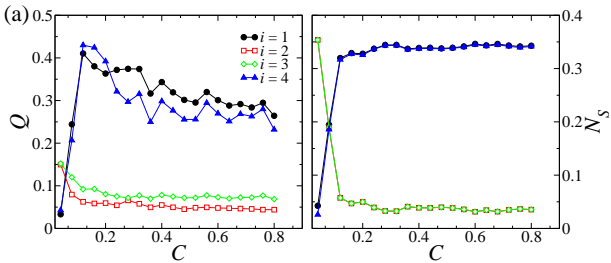}
\end{center}
\caption{Linear response $Q$ and normalized spike number $N_s$ versus activator coupling $C$. The system is composed by four units, and the other
parameters are: $\varepsilon = 10^{-4}$, $a_{1,4} = 1.01$, $a_{2,3} =
0.99$, $A_s = 0.01$, $D =0.50$, $T_s = 2.61$ and $\sigma_a^2 = 2 \cdot 10^{-7}$.} \label{fig:4QonCwc}
\end{figure}

\begin{figure}[ht!]
\begin{center}
\includegraphics[width=0.45\textwidth]{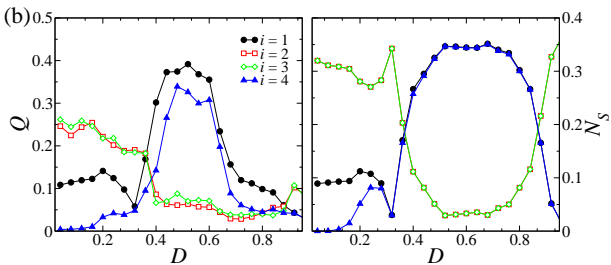}
\end{center}
\caption{Linear response $Q$ and normalized spike number $N_s$ versus inhibitory coupling $D$. The parameters are: $\varepsilon = 10^{-4}$, $a_{1,4} = 1.01$, $a_{2,3} = 0.99$, $A_s = 0.01$, $C =0.20$, $T_s = 2.61$ and $\sigma_a^2 = 2 \cdot 10^{-7}$.} \label{fig:4QonDwc}
\end{figure}

Figs.~\ref{fig:4QonCwc} and \ref{fig:4QonDwc} demonstrate that there are optimal values of couplings for the suppression to occur. While the dependence on the activator coupling $C$ is such that the suppression
holds for couplings larger than a given value, we observe a much narrower range, a resonance-like behavior,
as a function of the inhibitor coupling $D$. Even further, for $D$ large enough, the oscillation suppression
phenomenon disappears, and most of the spikes occur at frequencies different to the driving one (i.e. $Q$
vanishes).

\begin{figure}[ht!]
\begin{center}
 \includegraphics[width=0.45\textwidth]{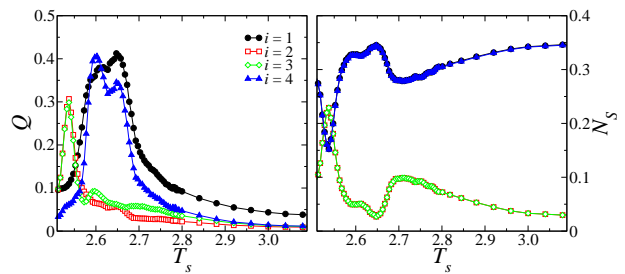}
\end{center}
\caption{Linear response $Q$ and normalized spike number $N_s$ versus driving period $T_s$, For this four units system, the parameters are: $\varepsilon = 10^{-4}$, $a_{1,4} = 1.01$, $a_{2,3} =
0.99$, $A_s = 0.01$, $C =0.20$, $D =0.50$ and $\sigma_a^2 = 2 \cdot 10^{-7}$.} \label{fig:4QonTwc}
\end{figure}

Fig.~\ref{fig:4QonTwc} shows the dependence on the signal periodicity $T_s$. It is
clear that the oscillation suppression and signal transmission are optimal
at the same parameter values. Furthermore, in the same figure it can be seen that there is a very narrow peak around the natural period ($T_s = T_{nat} = 2.54$) of the oscillatory units at which they
respond optimally. Note that the oscillation suppression holds for a wide range of
values of the driving period $T_s$. But the main result shown in this figure is the fact that the
suppression of oscillations in the oscillatory units is much more robust than in the
previous cases, i.e. $N=3$ and $N=4$ with strong coupling. This result is somewhat unexpected
given the fact that the couplings are not as strong as in the previous parameter sets, and then the units are allowed to move more freely.

\begin{figure}[ht!]
\begin{center}
\includegraphics[width=0.45\textwidth]{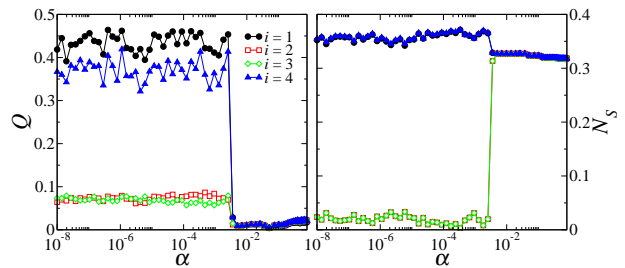}
\end{center} 
 \caption{The influence of the type of coupling on the the linear response $Q$ and the normalized
spike number $N_s$  in a system composed by four coupled FitzHugh--Nagumo's. The sliding parameter $\alpha$ shifts the weight of the
diffusion constant $D$ from a pure inhibitory coupling ($\alpha =
0.0$) to a pure activator coupling ($\alpha = 1$). The other parameters are
$\varepsilon = 10^{-4}$, $a_{1,4} = 1.01$, $a_{2,3} = 0.99$, $\sigma_a^2 = 2 \cdot 10^{-6}$, $A_s = 0.01$, $T_{s} = 2.61$, $C = 0.20$ and $D = 0.50$.}
\label{fig:4percentage}
\end{figure}

To show the different influence of activator and inhibitory couplings, we have added an extra activator
coupling in the model by interchanging $D$ by $(1-\alpha)D$ in the equations for the inhibitory variable
$y_i$ and inserting the terms $\alpha D(x_2-x_1)$ in Eq.~\ref{eq:x41}, $\alpha D(x_1-x_2)$ in
Eq.~\ref{eq:x42}, $\alpha D(x_4-x_3)$ in Eq.~\ref{eq:x43}, $\alpha D(x_3-x_4)$ in Eq.~\ref{eq:x44}. These
extensions of the model are used only in this section for the calculation of Fig.~\ref{fig:4percentage}. We
shift the balance between the activator and inhibitory coupling between these elements continuously with the
parameter $\alpha$. As shown in Fig.~\ref{fig:4percentage}, a very sharp transition to a situation of
oscillation suppression and no response to the driving frequency in the middle units is observed when the
activator coupling is strong enough, $\alpha > 3 \cdot 10^{-3}$).  Fig.~\ref{fig:4percentage} (as 
Fig.~\ref{fig:3percentage} for the $N=3$ case) shows the essential imperative of the inhibitory coupling
between the excitable and oscillatory units to reach the dynamic trap regime with the desired feature of
oscillation suppression and information transmission.

\section{Conclusions}

By studying coupled FitzHugh--Nagumo's units (or clusters, when one FitzHugh--Nagumo model represents many
similar neurons), we have found new mechanisms to suppress undesirable self-sustained oscillations in neuron
networks with a reconstruction of the desired functionality to submit information with the help of a
constructive role of noise. We have shown that an inhibitory coupling between oscillatory (malfunction) and
excitable neurons is essential to reach the dynamic trap regime which is responsible for the oscillation
suppression and the information transport. The dynamic trap regime is characterized by an anti-phase spiking
behavior of the excitable neurons at the end of the chain with the signal frequency and a silent
(oscillation suppressed) behavior of the originally oscillatory units in between. The desired dynamic trap
regime is sensitive with respect to the driving frequency, the noise intensity and the coupling strength.

Besides the dynamic trap regime we found other attractors which are similar to the dynamic trap attractor and offer also a reliable oscillation suppression but do not provide a good information transport along the chain. The oscillation suppression itself could be found in a larger parameter range
than the combination of oscillation suppression and reliable information transport.

It is interesting to note that the suppression is present not only due to noise but also in presence of a  strong enough driving force. It would then be interesting to study the effect of an alternative source of suppression of oscillations by the injection of an external signal.

In the present work we have considered chains of three or four units to keep the number of parameters small. Our model could be exemplary also for larger systems, if one regards one oscillator in the model as a representation of a cluster of many neurons in a close area with similar properties.

We have considered only paradigmatic models in a very general framework, hence we expect that these results
are also relevant to other  models with inhibitory coupling, which are used, for example, to describe various
physical \cite{1990_Kerner}, electronic \cite{1999_Ruwisch},  chemical  \cite{2000_Vanag,1990_Castets_prl}
systems, biological systems with spatial non uniformities  \cite{2003_Balazsi_chaos},  animal coat pattern
formation \cite{1982_Meinhardt}, or artificial gene networks synchronization with slow auto-inducer
diffusion \cite{2002_McMillen_pnas,2004_Kuznetsov}.

{\bf Acknowledgments.} AZ acknowledges financial support from Volkswagen-Stiftung (Germany). CJT wants to thank to the the Arbeitsgruppe Nichtlineare Dynamik at Universit\"at Potsdam for the kind hospitality extended to him. We acknowledge financial support from the Spanish Government and FEDER (EU) through projects FIS2004-5073, FIS2004-953, and the EU NoE BioSim (LSHB-CT-2004-005137).
 
\bibliography{refs}

\end{document}